\documentclass[a4paper,aps,prl,showpacs,preprintnumbers,twocolumn,amsmath,amssymb]{revtex4}

\usepackage{amsmath}
\usepackage{verbatim}
\usepackage{graphicx}
\usepackage{epsfig}

\newcommand{\one}{\mbox{$1 \hspace{-1.0mm}  {\bf l}$}}

\def\<{\langle}
\def\>{\rangle}
\newcommand{\Hc}{{\cal H}}
\newcommand{\Ac}{{\cal A}}
\newcommand{\Bc}{{\cal B}}
\newcommand{\Tc}{{\cal T}}
\newcommand{\be}{\begin{equation}}
\newcommand{\ee}{\end{equation}}
\newcommand{\bea}{\begin{eqnarray}}
\newcommand{\eea}{\end{eqnarray}}

\begin{document}

\title{Sequential generation of entangled multi-qubit states}

\author{C. Sch\"on,$^{1}$ E. Solano,$^{1,2}$ F. Verstraete,$^{1,3}$ J. I. Cirac$^{1}$, and M. M. Wolf,$^{1}$}

\affiliation{$^{1}$Max-Planck-Institut f\"ur Quantenoptik,
Hans-Kopfermann-Str. 1, D-85748 Garching, Germany \\
$^{2}$Secci\'on F\'{\i}sica, Departamento de Ciencias, Pontificia
Universidad Cat\'olica del Per\'u, Apartado Postal 1761, Lima,
Peru\\ $^{3}$Institute for Quantum Information, California
Institute of Technology, CA 91125, USA}

\pacs{03.67.-a,42.50.-p,03.65.Ud}
\date{\today}

\begin{abstract}
We consider the deterministic generation of entangled multi-qubit
states by the sequential coupling of an ancillary system to
initially uncorrelated qubits. We characterize all achievable
states in terms of classes of matrix product states and give a
recipe for the generation on demand of any multi-qubit state. The
proposed methods are suitable for any sequential
generation-scheme, though we focus on streams of single photon
time-bin qubits emitted by an atom coupled to an optical cavity.
We show, in particular, how to generate familiar quantum
information states such as W, GHZ, and cluster states, within such
a framework.
\end{abstract}

\maketitle

Entangled multi-qubit states are a valuable resource for the
implementation of quantum computation and quantum communication
protocols, like distributed quantum computing~\cite{DQC}, quantum
cryptography~\cite{Ekert1991} or quantum secret sharing
\cite{CGL99}. Using photonic degrees of freedom as qubits, say,
polarization states or time-bins of energy eigenstates, has the
advantage that photons propagate safely over long distances.
Consequently, photonic devices are the most promising systems for
quantum communication tasks. For this purpose, a lot of effort has
been made in recent years to develop efficient and deterministic
single photon sources~\cite{Law1997,Kuhn2002,McKeever2004,
Lange2004,Deppe2004, Forchel2004}.

Photonic multi-qubit states can be generated by letting a source
emit photonic qubits in a sequential manner \cite{Gheri2000}. If
we do not initialize the source after each step, the created
qubits will be in general entangled. Moreover, if we allow for
specific operations inside the source before each photon emission,
we will be able to create different multi-qubit states at the
output. In fact, this is a particular instance of a general
sequential generation scheme, where an ancillary system is coupled
in turn to a number of initially uncorrelated qubits.

It is the purpose of this paper to provide a complete
characterization of all multipartite quantum states achievable
within a sequential generation scheme. It turns out that the
classes of states attainable with increasing resources are exactly
given by the hierarchy of so-called matrix-product states
(MPS)~\cite{Zittartz,Fannes92}. These states typically appear in
the theory of one-dimensional spin systems \cite{AKLT}, as they
are the variational set over which Density Matrix Renormalization
Group techniques are carried out~\cite{DMRG}. Thus, our analysis
stresses the importance of MPS, since we show that they naturally
appear in a completely different and relevant physical context.
Moreover, particular instances of low-dimensional MPS, like
cluster states~\cite{Briegel2001} or GHZ states~\cite{GHZ}, are a
valuable resource in quantum information \cite{VerstCirQC}.
Conversely, we will provide a recipe for the generation on demand
of any multi-qubit state within a sequential generation scheme.
Due to the general validity of these results, we will first state
and prove them without referring to any particular physical
system. This will be then applicable to all sequential setups,
like streams of photonic qubits emitted either by a cavity QED
(CQED) source~\cite{Law1997,Kuhn2002,McKeever2004, Lange2004} or
by a quantum dot coupled to a microcavity~\cite{Deppe2004,
Forchel2004}.

In the second part, we will focus on the physical implementation
of these ideas within the realm of CQED. The role of the ancillary
system will be performed by a $D$-level atom coupled to a single
mode of an optical cavity.  The sequentially generated qubits will
be time-bin qubits $|0\>$ and $|1\>$, describing the absence and
presence of a photon emitted from the cavity in a certain time
interval (see Fig.\;1).

\vspace*{0.2cm}

\begin{figure}[h]
\begin{center}
\epsfig{file=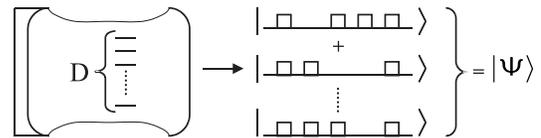,angle=0,width=0.8\linewidth}
\vspace*{-0.4cm}
\end{center}
\caption{A trapped D-level atom is coupled to a cavity qubit,
determined by the energy eigenstates $| 0 \rangle$ and $| 1
\rangle$. After arbitrary bipartite source-qubit operations,
photonic time-bins are sequentially and coherently emitted at the
cavity output, creating a desired entangled multi-qubit stream.}
\label{figure}
\end{figure}

We will concentrate on setups where all intermediate operations
are arbitrary unitaries and the ancilla decouples in the last
step. The latter enables us to generate pure entangled states in a
deterministic manner without the need of measurements.  Let
$\Hc_\Ac\simeq\mathbb{C}^D$ and $\Hc_\Bc\simeq\mathbb{C}^2$ be the
Hilbert spaces characterizing a $D$-dimensional ancillary system
and a single qubit (e.g. a time-bin qubit) respectively. In every
step of the sequential generation of a multi-qubit state, we
consider a unitary time evolution of the joint system
$\Hc_\Ac\otimes\Hc_\Bc$. Assuming that each qubit is initially in
the state $|0\>$ (i.e., the time-bin is empty), we disregard the
qubit at the input and write the evolution in the form of an
isometry $V:\Hc_\Ac\rightarrow\Hc_\Ac\otimes\Hc_\Bc$. Expressing
the latter in terms of a basis
$V=\sum_{i,\alpha,\beta}V_{\alpha,\beta}^i |\alpha, i \>\<\beta|$,
the isometry condition reads $\sum_{i=0}^1 V^{i\dagger} V^i
=\one$, where each $V^i$ is a $D \times D$ matrix. We choose a
basis where $\{ | \alpha \rangle , | \beta \rangle \}$ are any of
the $D$ ancillary levels. If we now apply successively $n$, not
necessarily identical, operations of this form to an initial state
$|\varphi_I\>\in\Hc_\Ac$, we obtain the state \be |\Psi\> =
V_{[n]} \ldots V_{[2]}V_{[1]}|\varphi_I\>\;. \ee Here, and in the
following, indices in squared brackets represent the steps in the
generation sequence.  The $n$ generated qubits are in general
entangled with the ancilla as well as among themselves. Assuming
that in the last step the ancilla decouples from the system, such
that $|\Psi\>=|\varphi_F\>\otimes|\psi\>$, we are left with the
$n$-qubit state \be\label{MPSiso} |\psi\>=\sum_{i_1\ldots
i_n=0}^1\<\varphi_F|V_{[n]}^{i_n}\ldots
V_{[1]}^{i_1}|\varphi_I\>\;|i_n,\ldots, i_1\> . \ee States of this
form are called matrix-product states
(MPS)~\cite{Zittartz,Fannes92}, and play a crucial role in the
theory of one-dimensional spin systems. Equation~(\ref{MPSiso})
shows that all sequentially generated multi-qubit states, arising
from a $D$-dimensional  ancillary system $\Hc_\Ac$, are instances
of MPS with $D \times D$ matrices $V^i$ and open boundary
conditions specified by $|\varphi_I\>$ and $|\varphi_F\>$. We will
now prove that the converse is also true, i.e., that every MPS of
the form \be\label{MPS} |\tilde{\psi}\> = \<\tilde{\varphi}_F|
\tilde{V}_{[n]}\ldots \tilde{V}_{[1]}|\tilde{\varphi}_I\>,\ee with
arbitrary maps
$\tilde{V}_{[k]}:\Hc_\Ac\rightarrow\Hc_\Ac\otimes\Hc_\Bc$, can be
generated by isometries of the same dimension, and such that the
ancillary system decouples in the last step. Since every state has
a MPS representation, this is at the same time a general recipe
for its sequential generation. The idea of the proof is an
explicit construction of all involved isometries by subsequent
application of singular value decompositions (SVD). We start by
writing $\<\tilde{\varphi}_F| \tilde{V}_{[n]}=V_{[n]}'M_{[n]}$,
where the $2\times 2$ matrix $V_{[n]}'$ is the left unitary in the
SVD and $M_{[n]}$ is the remaining part. The recipe for
constructing the isometries is the induction \be
\big(M_{[k]}\otimes\one_2\big)\tilde{V}_{[k-1]}
=V_{[k-1]}'M_{[k-1]}\label{induction},\ee where the isometry
$V_{[k-1]}'$ is constructed from the SVD of the left hand side,
and $M_{[k-1]}$ is always chosen to be the remaining part. After
$n$ applications of Eq.~(\ref{induction}) in Eq.~(\ref{MPS}), from
left to right, we set $|\varphi_I\>=M_{[1]}|\tilde{\varphi}_I\>$,
producing \be |\tilde{\psi}\> = V_{[n]}'\ldots
V_{[1]}'|\varphi_I\>.\ee Simple rank considerations show that
$V_{[n-k]}'$ has dimension $2\min{[D,2^k]}\times
\min{[D,2^{k+1}]}$. In particular, every $V_{[k]}'$ could be
embedded into an isometry $V_{[k]}$ of dimension $2D \times D$.
Physically, this just means we would have redundant ancillary
levels that we need not to use. Finally, decoupling the ancilla in
the last step is guaranteed by the fact that, after the
application of $V_{[n-1]}$, merely two levels of $\Hc_\Ac$ are yet
occupied, and can be mapped entirely onto the system $\Hc_\Bc$.
This is precisely the action of $V_{[n]}$ through its embedded
unitary $V_{[n]}'$.

This proves the equivalence of two sets of $n$-qubit states, which
are described either as $D$-dimensional MPS with open boundary
conditions, or as states that are generated sequentially and
isometrically via a $D$-dimensional ancillary system which
decouples in the last step. Motivated by current cavity QED
setups, we will now provide a third equivalent characterization,
namely, a set of multi-qubit states that are sequentially
generated by a source consisting of a $2D$-level atom. In contrast
to the first sequential scheme, the latter will not require
arbitrary isometries.

Consider an atomic system with $D$ states $| a_i \rangle$ and $D$
states $| b_i \rangle$, so that
$\Hc_\Ac=\Hc_a\oplus\Hc_b\simeq\mathbb{C}^d\otimes\mathbb{C}^2$.
That is, we will write $|\varphi\>|1\>$ for a superposition of $|
a_i \rangle$ states, whereas $|\varphi\>|0\>$ corresponds to a
superposition of $| b_i \rangle$ states. Since the last qubit
marks the atomic level, whether it belongs to the $| a_i \rangle$
or to the $| b_i \rangle$ subspace, we will refer to it as the
tag-qubit and write $\Hc_\Ac=\Hc_{\Ac'}\otimes\Hc_\Tc$. Now
consider the atomic transitions from each $| a_i \rangle$ state to
its respective $| b_i \rangle$ state accompanied by the generation
of a photon in a certain time-bin. This is described by a unitary
evolution, since now called ``D-standard map'', of the form  \bea
T:\;|\varphi\>_{\Ac'}|1\>_\Tc |0\>_\Bc &\mapsto&
|\varphi\>_{\Ac'}|0\>_\Tc |1\>_\Bc\;, \nonumber
\\ |\varphi\>_{\Ac'}|0\>_\Tc |0\>_\Bc &\mapsto&
|\varphi\>_{\Ac'}|0\>_\Tc |0\>_\Bc\; . \eea  Hence, $T$
effectively interchanges the tag-qubit with the time-bin qubit.
If, additionally, arbitrary atomic unitaries $U_{\Ac}$ are allowed
at any time, we can exploit the swap caused by $T$ in order to
generate the operation \be V|\varphi\>=\<0|_\Tc
T\Big(U_{\Ac}\big(|\varphi\>_{\Ac'}|0\>_\Tc\big)|0\>_\Bc\Big)
\label{IsofromT}\;,\ee which is the most general isometry
$V:\Hc_{\Ac'}\rightarrow \Hc_{\Ac'}\otimes\Hc_\Bc$. Therefore, the
so generated $n$-qubit states include all possible states arising
from subsequent applications of $2D \times D$-dimensional
isometries. On the other hand, they are a subset of the MPS in
Eq.~(\ref{MPS}) with arbitrary $2D \times D$-dimensional maps,
assuming that the atom decouples at the end. Hence, these three
sets are all equivalent.

Now, we show how these results can be applied in the realm of
cavity QED, where an atom is trapped inside a high-$Q$ optical
cavity, and we aim at generating multi-photon entangled states. A
laser may excite the atom, producing subsequently a photon in the
cavity mode, which, after some time, is emitted outside the cavity
(Fig.\ 1). We consider two different scenarios, corresponding to
the two families of states considered above. First, we may have
fast and complete access to the atom-cavity system. In
consequence, after the implementation of the desired isometry in
each step, we should wait until the photon leaks out of the cavity
before starting the next step. In this case, according to the
analysis above, we will be able to produce arbitrary
$D$-dimensional MPS with $D$ equal to the number of involved
atomic levels. Second, we may have a 2$D$--level atom ($D$ $| a_i
\rangle$ levels and $D$ $| b_i \rangle$ levels) and two kind of
operations: (i) fast arbitrary operations which allow us to
manipulate at will all atomic levels; (ii) an operation which maps
each $| a_i \rangle$ state to its corresponding $| b_i \rangle$
state while creating a single cavity photon, allowing a taylored
output. Here, we will also be able to produce arbitrary
$D$-dimensional MPS.

In the following, we will illustrate the above statements with a
specific example which is based on present cavity QED experiments
\cite{Kuhn2002,McKeever2004, Lange2004}. We consider a
three--level atom coupled to a single cavity mode in the
strong-coupling regime. An external laser field drives the
transition from level $|a\>$ to the upper level $|u\>$ with
coupling strength $\Omega_0$, and the cavity mode drives the
transition between $|u\>$ and level $|b\>$ with coupling strength
$g$, in a typical $\Lambda$ configuration [see Fig.\ 2(a)]. We
choose the detunings $\Delta$, with $|\Delta| \gg \{ g, \Omega_0
\}$, and assume that the cavity decay rate $\kappa$ is smaller
than any other frequency in the problem, so that we can ignore
cavity damping during the atom-cavity manipulations. By
eliminating level $|u\rangle$, we remain with an effective $D=2$
atomic system plus the cavity mode. We will show how, by
controlling the laser frequency and intensity, it is possible to
generate arbitrary $2$-dimensional MPS. Note that, by allowing the
manipulation of $D$ effective atomic levels, it is straightforward
to extend these results to the generation of $D$-dimensional MPS.

According to the results presented above, we just have to show
that we can implement any isometry
$V:\Hc_\Ac\rightarrow\Hc_\Ac\otimes\Hc_\Bc$. In fact, we will show
how it is possible to implement arbitrary operations on the atomic
qubit and the $\sqrt{SWAP}$ operation on the atom-cavity system,
which suffice to generate any isometry $V$ (since they give rise
to a universal set of gates for the two qubit
system~\cite{Nielsen2000}). The atomic qubit can be manipulated at
will using a Raman laser system, as it is normally done with
trapped ions~\cite{NIST,Innsbruck}. In order to implement the
$\sqrt{SWAP}$, we notice that the atom-cavity coupling is
described in terms of the Jaynes--Cummings model [see Fig.~2(b)],
where the coupling constant $\Omega_0$ is controlled by the laser.
Thus, application of laser pulses with the appropriate duration
and phase \cite{Innsbruck,SelectiveCQED} will implement the
unitary operation $U=e^{-iG}$, where generator
$G=(|a,0\rangle\langle b,1| + {\rm H.c.})\pi/4$, which corresponds
to the desired $\sqrt{SWAP}$ operation. In order to generalize
this scheme to an arbitrary $D$-level system, we notice that we
can view the atom as a set of $M$ qubits (with $D\le 2^M$). Thus,
if we are able to perform arbitrary atomic operations, together
with the $\sqrt{SWAP}$ operation on two specific atomic levels as
explained above, we can then implement a universal set of gates
and, in consequence, any arbitrary isometry.

\begin{figure}[h]
\begin{center}
\epsfig{file=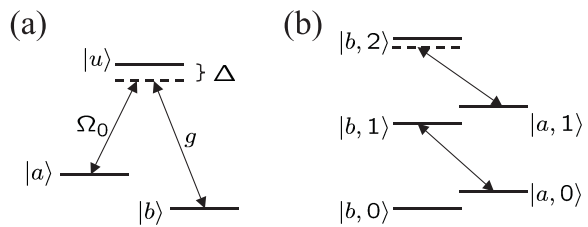,angle=0,width=0.9\linewidth}
\end{center}
\vspace*{-0.5cm}\caption {(a) Atomic level structure: levels
$|a\rangle$ ($|b\rangle$) and $|u\rangle$ are coupled by a laser
(cavity mode) off resonance. (b) After adiabatic elimination of
the upper state $| u \rangle$, we are left with a Jaynes--Cummings
type of Hamiltonian, where states $|a,n\rangle$ and
$|b,n+1\rangle$ are coupled. Both, the energy difference of those
levels and the corresponding Rabi frequency depends on $n$. The
reason for the first is the AC--Stark shift, whereas the second is
due to the Jaynes--Cummings coupling.} \label{figure2}
\end{figure}

In the rest of the paper, we will use another setup which is
closely related to current experiments
\cite{Kuhn2002,McKeever2004, Lange2004} and optimizes our second
method for MPS generation. In this frame, we will show how to
generate familiar multi-qubit states like $W$~\cite{Wstates},
$GHZ$~\cite{GHZ}, and cluster states~\cite{Briegel2001}, which are
all MPS with $D=2$ \cite{VerstCirQC}.

For the purpose, we consider an atom with three effective levels
$\{|a\>,|b_1\>,|b_2\>\}$ trapped inside an optical cavity. With
the help of a laser beam, state $|a\rangle$ is mapped to the
internal state $|b_1\rangle$, and a photon is generated, whereas
the other states remain unchanged. This physical process is
described by the map
\begin{eqnarray}
M_{\Ac\Bc}: \; | a \>  &\mapsto& | b_1 \> | 1 \> \; , \nonumber\\
| b_1 \> &\mapsto& | b_1 \> |0\> \; , \nonumber\\ | b_2 \>
&\mapsto& | b_2 \> |0\> \; ,
\end{eqnarray}
and can be realized with the techniques used
in~\cite{Kuhn2002,McKeever2004,Lange2004}. After the application
of this process, an arbitrary operation is applied to the atom,
which can be performed by using Raman lasers. The photonic states
that are generated after several applications are those MPS where
the isometries are given by $V_{[i]}=M_{\Ac\Bc}U_\Ac^{[i]}$, with
${i=1,\dots,n}$, $U_\Ac^{[i]}$ being arbitrary unitary atomic
operators.

For example, to generate a W-type state of the form
 \bea
  | \psi_{\rm W} \>
 &=& e^{i\Phi_1}\sin\Theta_1 |0...01\> + \cos\Theta_1
 e^{i\Phi_2}\sin\Theta_2 |0...010\> \nonumber\\&& +...+
 \cos\Theta_1...\cos\Theta_{n-2} e^{i\Phi_{n-1}}\sin\Theta_{n-1}
 |010...0\>\nonumber\\&& + \cos\Theta_1...\cos\Theta_{n-1}
 |10...0\>,
 \eea
we choose the initial atomic state $|\varphi_I\>=| b_2 \>$ and
operations $U_\Ac^{[i]} = U_{a {b_2}}^{b_1}(\Phi_i,\Theta_i)$,
with $i=1,\dots,n-1$, where \bea U_{kl}^m(\Phi_i,\Theta_i) &
\!\!\!\!\! = \!\!\!\! & \cos\Theta_i|k\>\<k| + \cos\Theta_i
|l\>\<l|+ e^{i\Phi_i} \! \sin\Theta_i|k\>\<l| \nonumber\\&& -
e^{-i\Phi_i}\sin\Theta_i|l\>\<k| + |m\>\<m| ,\eea and $\{ k,l,m \}
= \{ a,b_1,b_2 \}$. To decouple the atom from the photon state, we
choose the last atomic operation $U_\Ac^{[n]} = U_{a
{b_2}}^{b_1}(0,\pi/2)$ and, after the last map $M_{\Ac\Bc}$, the
decoupled atom will be in state $|b_1\>$.

To produce a GHZ-type state in similar way, we choose
$|\varphi_I\>=|a\>$, $U_\Ac^{[1]}=U_{a
{b_2}}^{b_1}(\Phi_1,\Theta_1)$, $U_\Ac^{[i]} = U_{a
{b_1}}^{b_2}(0,\pi/2)$, with $i=2,\dots,n-1$, and $U_\Ac^{[n]} =
U_{{b_1} {b_2}}^a(0,\pi/2)U_{a {b_1}}^{b_2}(0,\pi/2)$.

For generating cluster states, we choose $|\varphi_I\>=|b_2\>$,
$U_\Ac^{[i]} = U_{a {b_2}}^{b_1}(\Phi_i,\Theta_i) U_{a
{b_1}}^{b_2}(0,\pi/2)$, with $i=1,\dots,n-1$, and $U_\Ac^{[n]} =
U_{a {b_1}}^{b_2}(\Phi_n,\Theta_n) U_{{b_1}{b_2}}^a(0,\pi/2) U_{a
{b_1}}^{b_2}(0,\pi/2)$, obtaining \be |\psi \> =
\bigotimes_{i=1}^n \left(O_{i-1}^0 |0\>_i + O_{i-1}^1
|1\>_i\right) , \ee where $O^0_{i-1} = \cos\Theta_i |0\>_{i-1}\<0|
-e^{-i\Phi_i}\sin\Theta_i |1\>_{i-1}\<1|$ and $O^1_{i-1}=
e^{i\Phi_i}\sin\Theta_i |0\>_{i-1}\<0| + \cos\Theta_i
|1\>_{i-1}\<1|$, with ${i=2,\dots,n-1}$. Operators $O_{i-1}^0$ and
$O_{i-1}^1$ act on the nearest neighbor-qubit $i-1$ under the
assumption $O_0^0 \equiv \cos\Theta_1$ and $O_0^1 \equiv
e^{i\Phi_1}\sin\Theta_1$. If one chooses $\Phi_i=0$ and
$\Theta_i=\pi/4$ this leads to the cluster states defined by \be
|\psi_{\rm cl}\> = \frac{1}{2^{n/2}} \bigotimes_{i=1}^n
\left(\sigma^z_{i-1}|0\>_i + |1\>_i\right) \,\, , \,\, {\rm with}
\,\,\, \sigma^z_0\equiv 1 . \ee

The formalism presented here is also valid for other types of
single-photon sources, in the context of cavity QED or quantum
dots. For example, it could be extended to characterize the
polarization-entangled multi-qubit photon states generated by an
analogous cavity QED photon source~\cite{Gheri2000}. In fact, the
presented ideas and proofs apply to any multi-qudit state with
$\Hc_{\Bc}\simeq\mathbb{C}^d$ that is generated sequentially by a
$D$-dimensional source.

In a wider scope, we have established a formalism describing a
general sequential quantum factory, where the source is able to
perform arbitrary unitary source-qudit operations before each
qudit leaves. Apart from the multiphoton states, the present
formalism applies also to many other physical scenarios: (a) to
coherent microwave cavity QED experiments~\cite{Haroche}, where
atoms sequentially cross a cavity, and thus the outcoming atoms
end up in a MPS with the dimensions given by the effective number
of states used in the cavity mode; (b) a light pulse crossing
several atomic ensembles~\cite{Polzik2003}, which will be left in
a matrix product Gaussian state \cite{Norbert}; (c) trapped ion
experiments where each ion interacts sequentially with a
collective mode of the motion~\cite{NIST,Innsbruck,CiracZoller}.
Note also that one can include dissipation in the present
formalism, by replacing MPS by matrix product density operators
~\cite{Fannes92,MPDO}. This description applies, for example, to
the micromaser setup ~\cite{Walther2001} and other realistic
scenarios.

C.S. wants to thank K. Hammerer for useful discussions. This work
was supported by EU through RESQ project and the
"Kompetenznetzwerk Quanteninformationsverarbeitung der Bayerischen
Staatsregierung".

\end{document}